\documentclass{IEEEtran}
\usepackage{graphicx}
\usepackage{multirow}
\usepackage{epstopdf}
\usepackage{subfigure}
\usepackage{amsmath}
\usepackage{url}
\usepackage{float}

\begin{document}
\title{Toward Intelligent Network Optimization in Wireless Networking: An Auto-learning Framework}
\author{Wenyu Zhang, Zhenjiang Zhang, Han-Chieh Chao, Mohsen Guizani
\thanks{This article is going appear in IEEE Wireless Communications.}
\thanks{W. Zhang and Z. Zhang are with the School of Electronic and Information Engineering, Key Laboratory of Communication and Information Systems, Beijing Municipal Commission of Education, Beijing Jiaotong University, Beijing, 100044, China.}
\thanks{S. Zeadally is with the College of Communication and Information, University of Kentucky, Lexington, KY 40506 USA}
\thanks{H.-C. Chao is with the Department of Electrical Engineering, National Dong Hwa University, Hualien 974, Taiwan.}
\thanks{M. Guizani is with the College of Engineering, University of Idaho, Moscow, ID 83844-1023 USA.}}

\maketitle
\begin{abstract}
In wireless communication systems (WCSs), the network optimization problems (NOPs) play an important role in maximizing system performances by setting appropriate network configurations. When dealing with NOPs by using conventional optimization methodologies, there exist the following three problems: human intervention, model invalid, and high computation complexity. As such, in this article we propose an auto-learning framework (ALF) to achieve intelligent and automatic network optimization by using machine learning (ML) techniques. We review the basic concepts of ML techniques, and propose their rudimentary employment models in WCSs, including automatic model construction, experience replay, efficient trial-and-error, RL-driven gaming, complexity reduction, and solution recommendation. We hope these proposals can provide new insights and motivations in future researches for dealing with NOPs in WCSs by using ML techniques.
\end{abstract}


\section{Introduction}
In wireless communication systems (WCSs), network optimization problems (NOPs) have been extensively studied to maximize system performances by setting appropriate network configurations settings \cite{Tse2009Fundamentals}. NOP contains a broad range of research aspects in wireless networking, typical applications include resource allocation and management, system parameter provision, task scheduling, and user QoS optimization. Fig. \ref{NOP_framework} shows the basic process of solving a NOP in WCSs, which includes the following four steps:

\begin{figure*}
\centering
\includegraphics[width=0.8\textwidth]{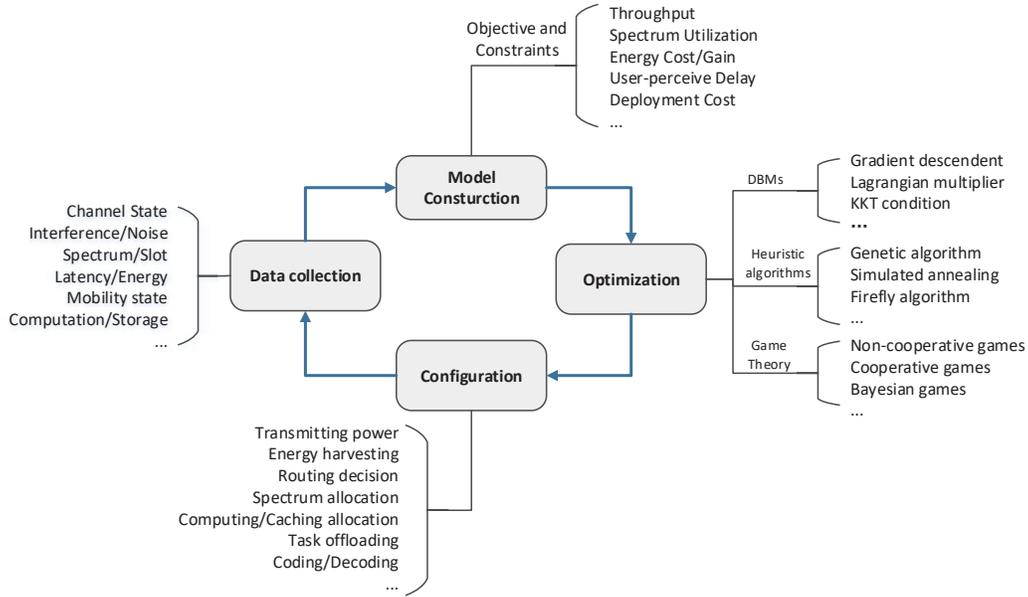}
\caption{Workflow of network management in wireless communication systems.}\label{NOP_framework}
\end{figure*}

\textbf{Data Collection}: which collects essential information of the system and the surrounding environment. The collected data can be channel state information (CSI), interference, noise, user location, spectrum and time-slot occupations, etc. Some QoS information, such as delay and energy consumption rates, mobility state, also can be the input data to support the following optimization process.

\textbf{Model Construction}: in which the expert constructs an optimization model that contains an objective function and several constraints. The objective of the optimization model can be throughput, spectrum utilization, user-perceive delay, energy consumption/gain, and facility deployment cost, etc. Typically, model construction is conducted by using a mathematical formulation process, and the experts are required to master the domain knowledge and theories involved in the model.

\textbf{Optimization}: The most commonly used methodologies for solving optimization problems are mathematical derivation-based methods (DBMs) and heuristic algorithms. The former one adopts a mathematical derivation process to find the solution, such as the Lagrangian multiplier, KKT conditions, and gradient descendent methodologies. The latter one adopts a heuristical neighborhood searching process to approach the optimal solution, including genetic algorithm, simulated annealing, particle swarm optimization, and firefly algorithms, etc. In general, DBMs are quite suitable for solving problems with explicit and convex objective functions, while heuristic algorithms does not require the derivatives of the objective functions, and are generally able to produce high-quality solutions for complex optimization problems if the optimization complexity is suitably high enough \cite{Chong2011An}. Except the above two optimization methods, game theoretical techniques, including non-cooperative games, cooperative games, and Bayesian games, also have been successfully applied to solve the optimization problem by learning automatic configuration strategies from the interactions with other functional nodes \cite{Saad2009Coalitional}.

\textbf{Configuration}: With the optimization results, the system then reconfigures the settings of the system to improve the performance. Possible reconfigurations may include transmission power allocation, energy harvesting scheduling, routing decision, spectrum resource allocation, to name a few. After configuration, the system then repeats the optimization process to keep the system in suitable working conditions.

Although NOPs have been extensively studied in WCSs, existing optimization methodologies still face the following three dilemmas:

\textbf{Human intervention}. The optimization models in NOPs are always constructed by experts with domain knowledge, and this knowledge-driven process is expensive and inefficient in practical implementations. If we can conduct the optimization operations automatically, network optimization will be more easy to be conducted in real world applications. However, how to reduce human intervention in solving NOPs is still a unexplored field in WCSs.

\textbf{Model invalid}. With the development of hardware and software techniques, the WCS is becoming an increasingly complex system with more users, more access ways, more complex functions and relationships among the network entities. In addition to transmitting power and the channel states, the system performances are also deeply influenced by the factors such as the software, hardware, interference, noise, and physical environment, and these factors are always unpredictable and hard to be formulated with explicit formulas. It is hard, or even unpractical for us to find the valid mathematical formulations for these factors, especially for the performance indices like delay and energy consumption rate influenced by the above unpredictable factors. In some situations, even if we have mathematical models to formulate the relationship functions, the actual implementation results are far from satisfactory due to the mismatches between theories and realities.

\textbf{High complexity}. Solving complex optimization problems may lead to expensive time cost due to the computation intensive optimization process, especially for complex NOPs with high dimensional solutions. In this situation, the efficiency of the algorithm may be unacceptable to meet the real-time requirement for delay-sensitive applications, such as gaming and vehicle networks. Even if the efficiency is acceptable, the continuous optimization process requires high computation energy cost in practical implementations. Predictably, in the future the complexity of WCSs will become more higher, and the corresponding NOP problem will also be more complex. Developing new effective and efficient models to solve these complex POPs is in urgent need in the research of future wireless networks.

In recent years, machine learning (ML) techniques have shown its powerful magics in dealing networking problems, such as traffic prediction, point-to-point regression, and signal detection \cite{Wang2017Machine}\cite{Chen2018Cognitive}. However, yet the application of ML in dealing NOPs has not been fully discussed in existing works. In this article, we focus on dealing with the NOPs in WCSs, and propose an auto-learning framework (ALF) that employs the ML techniques to achieve intelligent and automatic network optimization in WCSs. Within ALF, we propose several potential paradigms, including automatic model construction, experience replay, efficient trial-and-error, reinforcement learning-driven gaming, complexity reduction, and solution recommendation. The basic workflows, applications, and the challenges of these models will be discussed.

\begin{figure*}
\centering
\includegraphics[width=0.8\textwidth]{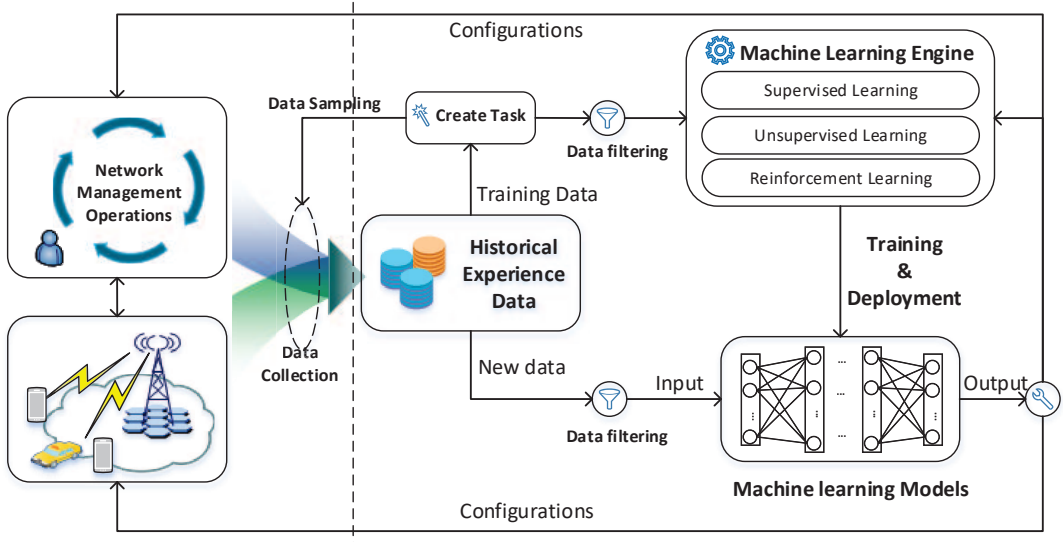}
\caption{Auto-learning framework for dealing with NOPs in WCSs.}\label{ALF_framework}
\end{figure*}

\section{Auto-learning Framework}
As shown in Fig. \ref{ALF_framework}, we propose an auto-learning framework (ALF) to achieve intelligent and automatic network optimization in WCSs. The basic workflow of ALF includes the following three steps:

\textbf{Data Collection}. Collecting the experience data is the prerequisite for conducting the ML-based models \cite{Chen2018From}, and must be properly addressed. Except the system and environment state information, in ALF the output solution data of an optimization process will also be collected as historical experience. When the training data are not sufficient, the system may need to conduct a resampling process to collect more data. A data filtering process needs to be done since the quality of used data has critical influences on the performance of the obtained black-box model. The outliers, incomplete data, and repeating data will be abandoned or refined in data filtering process.

\textbf{Model Training}. The model training process is conducted in a ML engine, in which different ML techniques are provided, including supervised learning, reinforcement learning (RL), and unsupervised learning. Their detailed application models will be introduced in the following section. After training, a cross validation process needs to be conducted to test the performance of the obtained model. More specifically, when the learning problem is a regression problem, i.e. outputs are continuous, the performance metric is the mean square error (MSE) between the predicted results and real outputs. When the outputs are discrete decisions, the problem can regarded as a classification problem, and the performance metric can be classification accuracy.

\textbf{Model Application}. Once a learning model is properly trained, it can be deployed in real-world WCSs. Given a new input instance, it passes through the mapping model and the corresponding output can be easily obtained in an efficient way.

$\bullet$ \textbf{Model Deployment}. The deployment of the mapping model is very easy to be achieved. The calculation process mainly includes matrix multiplications and non-linear transforms with activation functions, and both of them can be easily calculated.

$\bullet$ \textbf{Model Refinement}. The black-box auto-learning model may need to be refined due to the change of wireless systems and environments, and imperfect training data. The dynamic adjusting of a ML-model can be regarded an incremental learning problem, and the key step is the proper updating of training data instances. Therefore, it is suggested to updating the training data set periodically to guarantee the obtained model perform well when the system model is changed.

\section{Supervised Learning: Automatic Model Construction and Experience Replay}
With sufficient training data, a complex non-linear mapping function from input data space to the output data space can be obtained by training a supervised learning model. Benefit from this learning ability, supervised learning has been successfully applied in point-to-point learning tasks in communications systems, such as delay prediction, channel estimation and signal detection \cite{Chen2018Label}. According to the amount of training samples, supervised learning can be divided into the following two categories: small-sample learning (SSL) and deep learning (DL). Possible choices of SSL include shallow neural networks, kernel-based methods, and ensemble learning methods. For DL, possible choices include deep belief networks, and deep Boltzmann machines, and deep convolutional neural networks \cite{Abramson2006Pattern}.

\subsection{Automatic Model Construction}

\begin{figure*}
\centering
\includegraphics[width=0.75\textwidth]{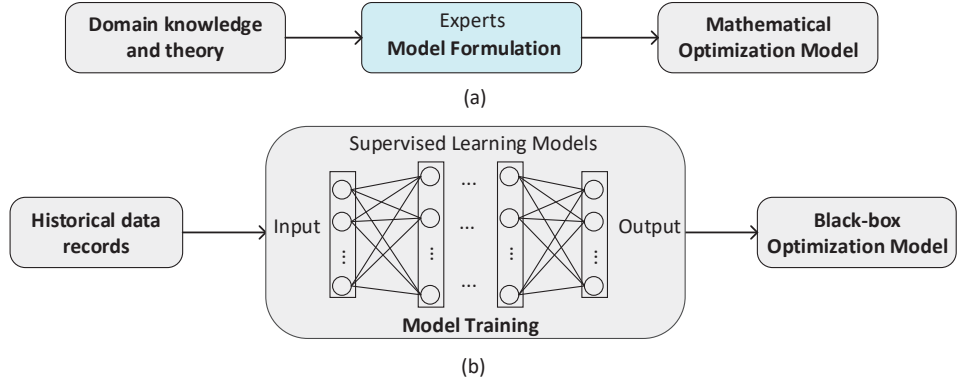}
\caption{Comparison of model construction process, in which (a) conventional mathematical model construction, and (b) automatic model construction using supervised learning-based regression techniques.}\label{AutoCons}
\end{figure*}

\textbf{Model}: Supervised learning-based black-box regression provides an effective way to solve the expensive human intervention and model invalid problems. In situations when the explicit functions between the input and output are not available, but we have sufficient data samples that contain the inputs and outputs of the system, the mapping function can be trained by using a supervised regression technique. Given a new input data, the target performance objective can be accurately predicted by using the previously obtained model.

We propose to use supervised learning techniques to automatically conduct the model constructions process in NOPs. As illustrated in Fig. \ref{AutoCons}(a), in conventional NOPs, the mathematical optimization model is constructed by experts with domain knowledge. In ALF, we propose to use black-box modeling to automatically construct the optimization model, as shown in Fig. \ref{AutoCons}(b). In the automatic model construction process, we can directly regress the objective function and constraints by using regression models. In the same way, the constraints can also be constructed. With the obtained model, a following heuristic algorithm can be used to solve the optimization model, since it just needs to know the objective response in each searching iteration.

When the target function contains several independent parts, we can firstly train the independent mapping functions of these parts, and then combine them into a unified one. For example, in mobile edge computing, the user-perceive delay mainly includes three parts: data transmission time, queuing time, and task execution time. In this scenario, we can build the optimization model by combining the three black-box delay time prediction models.

\textbf{Challenges}: The successful implementation of a supervised learning method requires a dataset with sufficient and reliable data samples to train the mapping model. In some tasks like network delay and energy consumption rate prediction, the data samples can be easily collected. However, collecting a large number of data samples in a short time may be unpractical for some systems with very high reconfiguration cost, such as the reconfiguration of virtualized network function resources in software defined WCSs. Therefore, how to reduce training data samples is critical in automatic model construction-based NOPs.

\begin{figure*}
 \centering
 \subfigure[]{
 \includegraphics[width=0.7\textwidth]{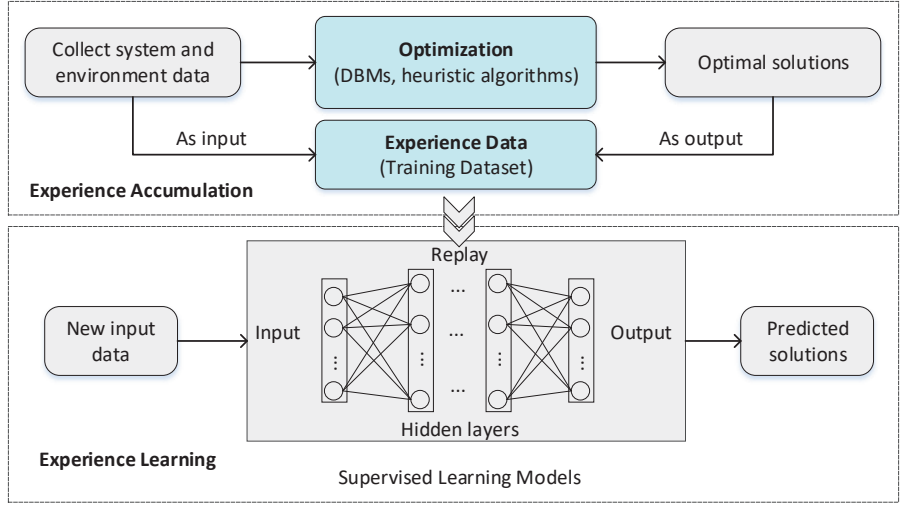}}
  \subfigure[]{
 \includegraphics[width=0.4\textwidth]{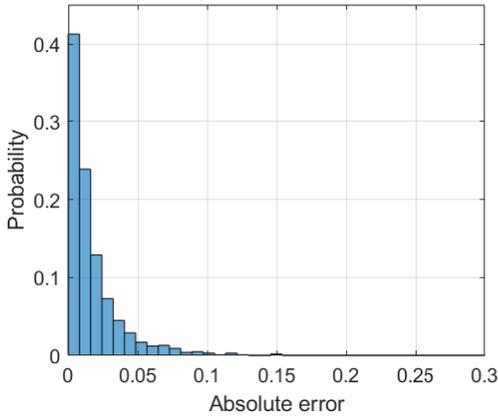}}
   \subfigure[]{
 \includegraphics[width=0.4\textwidth]{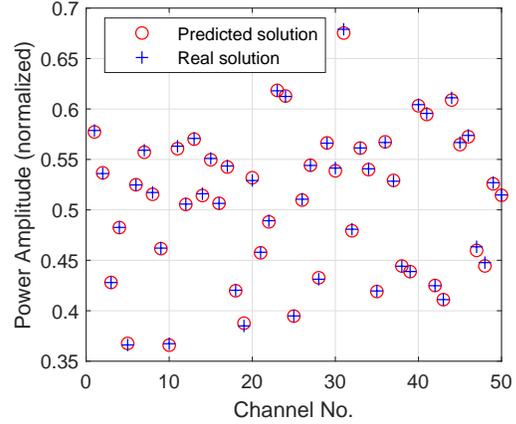}}
 \caption{Illustration of experience replay-based fast optimization, in which (a) shows the workfow, (b) and (c) provide an example of an power allocation in a massive MIMO system with 50 antennas.} \label{Experience_Replay}
\end{figure*}

\begin{table*}[!htbp]
  \centering
  \caption{Performance comparison of Self-dual embedding and experience replay}
    \begin{tabular}{|c|c|c|c|} \hline
         &  Training time (s)     &  Deployment time (s) &  Average absolute error \\ \hline
      Self-dual embedding    & --      &   1.21 & 0 \\ \hline
       Experience replay    &   $1.2\times 10^4$    &  $2.38\times 10^{-4}$  & 0.018  \\ \hline
    \end{tabular}%
  \label{tab:compa}%
\end{table*}%

\subsection{Experience Replay}
\textbf{Model}: For intelligent biological individuals, learning from their experiences is a common practice to improve the efficiencies of their behaviors. In conventional NOPs, although the system may repeatedly conduct the optimization process, the historical experiences are actually abandoned and can not be fully utilized. By exploiting supervised learning techniques, one can train a learning model that directly maps the input parameters to the optimization solutions. In this way, the repeating optimization process with high complexity can be avoided, the solution can be predicted with very low computation cost. The workflow of experience replay-based optimization is shown in Fig. \ref{Experience_Replay}(a), which includes the following two phases:

\textbf{Experience Accumulation}: When the optimization model is deployed in the network manager, its historical input data and the obtained optimal configurations can be used as the experience (or training data) to train a supervised learning model. To achieve this goal, firstly we construct the optimization model, then an optimization process is developed to find the optimal solutions. Given an input parameter data, its corresponding optimal solution that achieves best performance will be regarded as the output. The whole data collection process can be obtained by repeating the sampling or reconfiguration process, and it is terminated until we have sufficient data samples such that the prediction performance of the model is satisfied.

\textbf{Experience Learning}: A supervised learning-based solution prediction model can be properly trained with the obtained training data. Note that the used training data may need to be normalized before the training process. One can use an online model training process by using the gradient descendent process, or directly using the whole historical experience dataset to train the model offline. The choice of the learning model plays an important role in determining the model's prediction performance. Note that, although DL may have stronger generalization capacity when dealing with big data compared with SSL, it doesn't mean DL is always a better choice than SSL, because a proper training of DL model is much more expensive compared with SSL, and SSL always outperform DL when the data sample is small.

\textbf{Applications}: Experience replay can be trained both online and offline, and requires much less training data samples compared with reinforcement learning since the training data are all optimal results. In this way, many resource management applications, such as power allocation for OFDM and massive MIMO signals can be speed-up by using experience replay.

We test the performance of experience replay for power allocation to maximize the throughput of a Massive MIMO system with 50 antennas, the details of parameter settings and the used self-embedding baseline technique can be seen in \cite{Bjornson2013Massive}. In this test, the CSI data is used as input data, and the power allocations of antennas are output data. Subfigure \ref{Experience_Replay}(b) shows the probability density distribution of the absolute errors between the predicted results and the real solution, and subfigure \ref{Experience_Replay}(c) depicts an example of one channel realization. Table \ref{tab:compa} shows the average performance comparison results. In this test, the kernel extreme learning machine (KELM) is used as the learning model for its excellent performances on high regression precision and low computation efficiency \cite{Huang2011Extreme}.

We generate 10000 allocation experience instances by using the self-embedding technique, in which 9000 instances are randomly selected for training, and the remaining 1000 instances are used for testing. All data instances, including input data and output data, are normalized between interval [0,1]. We can see that the average absolute error between predicted results and real solutions is only 0.018, and results in subfigures (b) and (c) also show that prediction errors are very small. Most importantly, the deployment time of one optimization using self-embedding technique is $1.21s$, but experience replay only needs $2.38\times 10^{-4}s$ to predict the results, which is much more fast. Since we need to collect sufficient training data instances, the total time cost of training process is about $1.2\times 10^4s$. However, the training process is conducted before the algorithm is embedded in the system, thus it has no influences on practical implementation. These results prove that experience replay can significantly improve the efficiency of computation intensive network optimization operations, at the same time guarantee high-quality prediction performances.

\textbf{Challenges}: The successful implementation of experience replay relies on the high-quality experience data and the generalization capacity of the adopted learning algorithm. For NOPs without high-quality solutions, the obtained prediction model maybe biased due the imperfection of the training data. Moreover, when the dimension of the solution is very large, the model may be unable to be properly trained even when the adopted learning model has a strong regression ability. As a result, the predicted results may suffer from a performance loss compared with the results of conventional optimization results.

\section{Reinforcement Learning: Efficient Trial-and-Error and RL-driven Gaming}
Reinforcement learning (RL) can be used to learn the decision policies to automatically take actions to maximize the reward of the agent in a certain environment \cite{Chen2018From}. It is known that RL can be used to solve optimization problems without requiring objective functions and environment conditions \cite{He2018Deep}. On the other hand, Bayesian optimization method is an effective statistical inference learning-based optimization method for problems without explicit objective functions, and it have been proved valuable in providing efficient and effective frameworks to train ML models \cite{Snoek2012Practical}.

\subsection{Efficient Trial-and-Error}
\textbf{Model}: In RL-based decision making, the agent collects the system state and reward from the environment, and trains a Markov decision process to take actions according to the current environment state and reward. The policy map and environment transition probabilities are updated by using dynamical interactions with the environment. The most commonly used RL model is the Q-learning model, in which the manager intends to maximize the Q-value of the by using an iterative learning process, as given by
\begin{equation}\label{Qf}
Q^{*}(s,a) \leftarrow Q(s,a) + \alpha [R(s,a) + \gamma \max_{a\in \mathcal{A}} Q(s^{*},a^{*})-Q(s,a)],
\end{equation}
where $s$ and $a$ denote the state and the action of the system, respectively, $R(s,a)$ denotes the corresponding reward, $\mathcal{A}$ is the set contains all possible actions. Parameter $\alpha$ is the learning rate adjusting the convergence speed of the learning process, and parameter $\gamma$ controls the decaying speed of the impact of historical experience on the Q-value. In each iteration, the $\epsilon$-greedy selection strategy is usually used to decide whether accepting a better result, and $\epsilon$ denotes the acceptance probability.

When dealing with decision making problems in wireless networking, an alternative is training a neural network-based RL model to automatically to make decisions without any model of the target system, and usually this RL-model can be trained by using a policy gradient descent (PGD) method. However, the training process requires a large number of reconfiguration trials, which limits its application in wireless networking. By integrating Bayesian learning in RL, the PGD process can be replaced by an efficient trial-and-error optimization process to obtain the parameters of the RL model. In this way, the convergence process RL learning process can be more fast, and the number of reconfiguration trails will be greatly reduced.

\textbf{Applications}: Mobile edge computing (MEC) and fog computing provides low-latency computation and caching services for mobile user terminals \cite{Zhang2017Cooperative}. In MEC, some NOPs like content caching, task offloading scheduling, task assignment in the Cloudlet server can be achieved by using the efficient trial-and-error RL model. In addition, Bayesian optimization itself can be used to derive optimal system provision parameters in WCSs.

\subsection{RL-driven Gaming}
\textbf{Model}: Game theory has been a powerful tool in guiding the behaviors in interacting with other entities of the wireless network. In conventional game theoretical models, all users adopts a knowledge-based mathematical model to learn the optimal strategies to maximize their own benefits (or utilities) \cite{Saad2009Coalitional}. In general, when the users are rational, and know how to maximize their own rewards, a Nash equilibrium can be achieved by repeating the gaming process with proper strategy update processes. In a similar way, the game model can be solved by using a multi-task learning (MTL) framework with reinforcement learning techniques.

The implementation of RL-based MTL is quite similar to the strategy updating process in conventional game theoretical models. In each repetition, all the agents, or users, select their own actions and execute the selected action, then observe the new state of the system and reward obtained. Subsequently, the strategies are updated by using equation \eqref{Qf}. By repeating the above process, the Nash equilibrium can be achieved in the whole system. In addition, the above RL-based gaming approach can be improved by using a cooperative gaming process, in which the users is able to know the rewards of other users, but not just their own rewards. In this way, the repetition numbers can be reduced.

\textbf{Applications}: RL-driven gaming can be used in device-to-device (D2D) networks and cognitive radio (CR) networks. In D2D networks, the devices directly communicate with each other without the relay of base stations. RL-driven gaming can be used to design the communication choices of the devices to maximize their own performances. In CR, the secondary users want to maximize their own communication capacity, but can not interfere the communications of the primary users. RL-driven gaming can be used to design the spectrum occupying behaviors of both primary users and secondary users.

\textbf{Challenges}: The convergence of RL-driven gaming may require a large number reconfiguration trials, which is not efficient compared with conventional model based game models, thus using the efficient trial-and-error is also meaningful in this scenario. Reducing the sampling number as many as possible is still an open problem needs to be further studied.

\section{Unsupervised learning: Complexity Reduction and Solution Recommendation}
Clustering algorithm is one typical unsupervised learning method that aims at partitioning the data into several clusters with similar regional distribution properties. The $k$-means algorithm is an efficient and effective clustering algorithm, and it can be used to solve most clustering problems \cite{Witten2012Data}. Also, the similarity learning process used in $k$-nearest neighbor ($k$-NN) search can be used in finding recommended solutions.

\subsection{Complexity Reduction}

\begin{figure}
\centering
\includegraphics[width=0.5\textwidth]{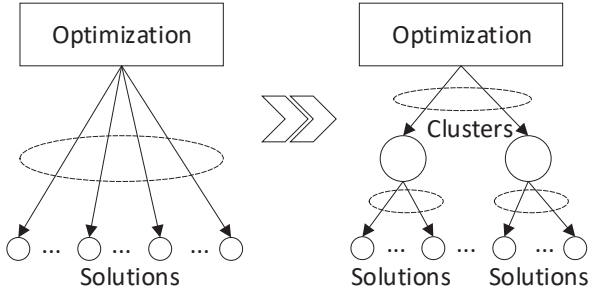}
\caption{Hierarchical optimization for complexity reduction.}\label{complexity_red}
\end{figure}

\textbf{Model}: It is recognized that the increasing of variable dimensions will greatly increase the complexity of optimization process. We therefore discuss the potentials of using clustering algorithms to reduce the complexity of NOPs with high-dimensional variables. As shown in Fig. \ref{complexity_red}, we can modify the original NOP into a hierarchical NOP problem to reduce the complexity by dividing the target high-dimensional variables into several clusters. Firstly, cluster-level optimization process is conducted, then variable-level optimization is executed within each clusters. In this way, since the cluster number and variable dimension of each cluster is much more smaller than original variable vector, the complexity of optimization process can be greatly reduced.

\textbf{Applications}: In applications like resource management with large number of variables, the optimization process can be an expensive task with high dimensional target variables. In this situation, the model complexity can be relaxed by using a clustering process. The variable vector can be divided into several sub-vectors according to factors like throughput demand, channel states, computation demands, and data transmission amount, etc. Some other factors, such as user priority, geographical position, and residual energy, also can be used as the features for clustering. By this way, optimization can be conducted in cluster level and task level separately, and the complexity can be significantly reduced.

\textbf{Challenges}: The drawback of clustering-based hierarchical optimization is that, the obtained results may suffer from a performance loss since the hierarchical optimization process is not the same to the original one, and cluster-optimal results are not equivalent to variable-optimal results. Therefore, How to reduce the performance loss in hierarchical optimization is the challenge of future's work.

\subsection{Solution Recommendation}

\textbf{Model}: One can use a similarity measurement to find similar historical tasks, then directly combine the solution of these similar task as the solution of the new task. To realize similarity-based solution recommendation (SSR) in ALF, firstly we define the feature vector that is able to distinguish the differences of the tasks, and subsequently a $k$-NN searching process can be used to find the tasks with similar features. The $k$-NN algorithm is a well-known lazy learning method that searches the nearest instances according to similarity measurements, and it can be efficiently realized by using a kd-tree algorithm. We assume that the environment keeps stable in a period of time. Given a new task, when the historical tasks with similar features are known, we can combine the solutions of these similar tasks, and directly use the average result as the solution.

\textbf{Applications}: Large scale power allocation is usually a computation intensive task due to the high dimension of the solution. If we have sufficient historical feature data, the SSR can be used to solve the real-time optimization problem. The feature data can be defined as a vector contains user geographic location and user terminal type. When the locations are close with each other, the corresponding CSI will be similar. In addition, when the user terminal type is the same, their antenna capacities will also be the same. In this way, the power assignments will also be similar.

\textbf{Challenges}: First, collecting user feature data may impose privacy concerns since the manager may want to collect sensitive information, such as geographic locations, user behaviors, and user preferences. Second, since SSR assumes that the environment keeps static in a period time, it is not able to deal with problems with dynamic, or stochastic conditions. Third, the recommended solution is just an approximated version of the real one, and the corresponding performance will also be not optimal. Forth, the distribution of the collected data may not be evenly distributed. For some new tasks without sufficient close neighbor, SSR will be failed to find the reliable results.

\section{Conclusions}
This article recalled the models of network optimization in WCSs and proposed an ALF that employs the advantages of powerful ML techniques to deal with the human intervention, model invalid, and high complexity problems in conventional optimization models. We reviewed the basic concepts of supervised learning, reinforcement learning, and unsupervised learning, and then proposed their several potential models to deal with NOPs, including automatic model construction, experience replay, efficient trial-and-error, RL-driven gaming, complexity reduction, and solution recommendation. We encourage the readers to test and modify these proposals, and further design more new ML-based methods for dealing with NOPs in WCSs.

\bibliographystyle{unsrt}
\bibliography{References}

\end{document}